\documentclass{llncs}
\usepackage{amsmath}
\usepackage{amssymb}
\usepackage{comment}
\usepackage{hyperref}
\usepackage{longtable}
\usepackage{stmaryrd}
\newcommand{\interp}[1]{\llbracket #1 \rrbracket}
\newcommand{\maps}{\colon}
\newcommand{\FinSet}{\mathrm{FinSet}}
\newcommand{\Set}{\mathrm{Set}}
\newcommand{\Cat}{\mathrm{Cat}}
\newcommand{\Calc}{\mathrm{Calc}}
\newcommand{\Mon}{\mathrm{Mon}}
\newcommand{\BoolAlg}{\mathrm{BoolAlg}}
\renewcommand{\Form}{\mathrm{Form}}
\newcommand{\leftu}{\mathrm{left}}
\newcommand{\rightu}{\mathrm{right}}
\newcommand{\send}{\mathrm{send}}
\newcommand{\recv}{\mathrm{recv}}
\newcommand{\comm}{\mathrm{comm}}
\renewcommand{\quote}[1]{``#1"}
\newcommand{\deref}[1]{\mathrm{eval}(#1)}
\newcommand{\op}{\mathrm{op}}
\newcommand{\NN}{\mathbb{N}}
\makeatletter
\gdef\tshortstack{\@ifnextchar[\@tshortstack{\@tshortstack[c]}}
\gdef\@tshortstack[#1]{%
  \leavevmode
  \vtop\bgroup
    \baselineskip-\p@\lineskip 3\p@
    \let\mb@l\hss\let\mb@r\hss
    \expandafter\let\csname mb@#1\endcsname\relax
    \let\\\@stackcr
    \@ishortstack}
\makeatother

\title{Logic as a distributive law}
\author{
Michael Stay\inst{1}\\
\and
L.G. Meredith\inst{2}\\
}
\institute{
  {Pyrofex Corp.}\\
  \email{\fontsize{8}{8}\selectfont stay@pyrofex.net}\\
  \and
  {Synereo, Ltd}\\
  \email{\fontsize{8}{8}\selectfont greg@synereo.com}
}
\begin{document}
\maketitle
\begin{abstract}
\noindent
  We present an algorithm for deriving a spatial-behavioral type
  system from a formal presentation of a computational calculus.
  Given a 2-monad $\Calc\maps \Cat \to \Cat$ for the free calculus on
  a category of terms and rewrites and a 2-monad BoolAlg for the free
  Boolean algebra on a category, we get a 2-monad Form = BoolAlg +
  Calc for the free category of formulae and proofs.  We also get the
  2-monad $\BoolAlg \circ \Calc$ for subsets of terms.  

  The interpretation of formulae is a natural transformation
  $\interp{-} \maps \Form \Rightarrow \BoolAlg \circ \Calc$ defined by
  the units and multiplications of the monads and a distributive law
  transformation $\delta\maps \Calc \circ \BoolAlg \Rightarrow
  \BoolAlg \circ \Calc.$  This interpretation is consistent both with
  the Curry-Howard isomorphism and with realizability.  

  We give an implementation of the ``possibly'' modal operator
  parametrized by a two-hole term context and show that, surprisingly,
  the arrow type constructor in the $\lambda$-calculus is a specific
  case.  We also exhibit nontrivial formulae encoding confinement and
  liveness properties for a reflective higher-order variant of the
  $\pi$-calculus.  

\end{abstract}
\section{Introduction}
  
  Sylvester coined the term ``universal algebra'' to describe the idea of expressing a mathematical structure as set equipped with functions satisfying equations; the idea itself was first developed by Hamilton and de Morgan \cite{Graetzer}.  Most modern programming languages have a notion of a ``module'' or an ``interface'' with this same information; for example, consider this Agda definition of a monoid.
  \begin{verbatim}
    data Monoid (M : Set)(Eq : Equivalence M) : Set where
      monoid :
        (e   : M)
        (_*_ : M -> M -> M)
        (leftId : LeftIdentity Eq z _+_)
        (rightId : RightIdentity Eq z _+_)
        (assoc : Associative Eq _+_) ->
           Monoid M Eq
  \end{verbatim}
  The code defines a sort $M$, a nullary term constructor $e$, and a
  binary term constructor $*,$ subject to three equations. In
  universal algebra, such a structure is called an ``equational
  theory''.
  
  In 1963, Lawvere \cite{Lawvere} showed that an equational theory was
  a presentation of a category with finite products where all the
  objects are powers of a single generating object; such a category is
  now called a Lawvere theory.  The Agda code above is a presentation
  of a category Th(Mon) whose objects are $1, M, M^2, M^3, \ldots,$
  and whose morphisms are generated by projections, $e,$ and $*,$
  modulo the equations.  This category is purely syntactic.

  To interpret the equational theory as denoting sets and functions,
  we use a product-preserving functor from the Lawvere theory to Set:
  such a functor maps $M$ to a set and $e$ and $*$ to functions
  satisfying the equations.  The category Set encodes what we mean by
  semantics; if instead of Set we used the category Vect of vector
  spaces and linear maps and mapped the product in Th(Mon) to the
  tensor product in Vect, a model of the theory would be an
  associative algebra instead.  If we used a category of endofunctors
  and natural transformations and mapped the product in Th(Mon) to
  composition, a model of the theory would be a monad.

  The category of product-preserving functors from Th(Mon) to Set and
  natural transformations between them is equivalent to the category
  Mon of monoids and monoid homomorphisms.  There is a forgetful
  functor $U\maps \Mon \to \Set$ that forgets all the structure with a
  left adjoint $L\maps \Set \to \Mon$ that picks out the free monoid
  on a set.  Abusing notation somewhat, we use Mon also to mean the
  monad $UL$ that picks out the underlying set of the free monoid on a
  set.  This pattern is a general one: every Lawvere theory
  corresponds to a finitary monad on Set and vice-versa
  \cite{DBLP:journals/entcs/HylandP07}.

  Often modules or interfaces are used to model data structures.  In
  computing, however, we consider not only the structure of data but
  the behavior of processes that change the data.  The
  $\lambda$-calculus is the paradigmatic model of functional
  programming.  It has a single data type, called lambda term,
  described by a one-line grammar.  In this model, computation
  consists of repeatedly applying a single rewrite rule called
  $\beta$-reduction; the rewrite rule matches the current term to a
  pattern, then rearranges the parts of the term.  When the term no
  longer matches the pattern, the rewriting stops and the resulting
  term is taken as the ``answer''.  Similarly, the $\pi$-calculus,
  arguably the paradigmatic model of concurrent programming, has a
  structured data type called a process that is also described by a
  simple grammar.  Computations in the $\pi$-calculus are carried out
  by repeatedly applying rewrite rules, primarily the
  $\mathsf{comm}$-rule.

  This pattern has been recognized and formalized many times.  In his
  seminal paper, ``Functions as Processes''
  \cite{DBLP:journals/mscs/Milner92}, Milner developed what is now the
  standard presentation of a computational calculus.  The presentation
  consists of a grammar describing the primary data type over which
  computations are carried out, a structural equivalence, used to
  erase syntactic differences that are irrelevant to computation, and
  a set of rewrite rules describing how to realize computation through
  operations on the data structures.  On one hand, this generalizes
  universal algebra's generators and relations presentation, with the
  grammar replacing the generators as the free construction, the
  structural equivalence replacing the relations, and the rewrites
  providing the computational semantics.  On the other hand, this
  presentation maps quite well onto Plotkin's SOS format for
  operational semantics \cite{Plotkin04theorigins}, but the scope is
  much larger.  Even the presentation of the Java VM can be seen as
  essentially a rewrite system with the state of the virtual machine
  as instances of a data type over which computation is carried out,
  and the transitions of the virtual machine as the rewrite rules
  \cite{DBLP:conf/oopsla/IgarashiPW99}.  Capturing this higher-order
  computational phenomena, expressed as generalized rewrites, is what
  motivates our movement to higher categorical semantics.

  Lawvere theories can be generalized from  1-categories to 2-categories.  As one might expect, the corresponding notion of equational theory involves one more level of structure; we have sorts, term constructors, {\em rewrites} instead of equations between term constructors, and equations between rewrites.  We typically interpret such a theory in Cat, the  2-category of categories, functors and natural transformations.

  We can use Lawvere 2-theories to describe terms in a computational calculus and the permissible rewrites.  We have a sort for terms;  our examples are necessarily simple calculi due to space constraints, but our approach should work with any formalization of a computational model such as the K framework \cite{DBLP:journals/jlp/RosuS10}, the  ``kiama'' Scala package \cite{DBLP:conf/gttse/Sloane09}, even packages largely focused on documentation and specification, such as Ott  \cite{DBLP:journals/jfp/SewellNOPRSS10}.  We also have term constructors for building up specific terms, rewrites between term constructors for running programs, and equations between rewrites to preserve invariants.  The Lawvere 2-theory captures the syntax of a computational calculus; a 2-functor from the theory into Cat assigns operational semantics.  Note that this is fully consistent with Curry-Howard style interpretations.

  A spatial-behavioral type system is a language in which one can describe both the structure and future behavior of terms.  In keeping with realizability style semantics  \cite{Krivine-TheCurryHowardCorre}, the interpretation of a formula should be a set of terms satisfying the formula.  The interpretation of a proof should be a function that maps a set of assumptions to a set of consequents.

  The obvious language for describing the structure of a term is the 2-theory Th(Calc) of the calculus itself, equipped with extra term constructors for true, false, disjunction, conjunction, and negation.  This suggests adding the 2-monad Calc to the 2-monad BoolAlg for Boolean algebras to get Form, the 2-monad for formulae.

  The obvious language for interpreting formulae is the 2-category of subsets of terms with pointwise rewrites between them.  This suggests composing the monad BoolAlg with the monad Calc.

  The obvious way to interpret formulae is then a natural transformation $\interp{-} \maps \Form \Rightarrow \BoolAlg \circ \Calc$ defined by the units and multiplications of the monads and a distributive law transformation $\delta\maps \Calc \circ \BoolAlg \Rightarrow \BoolAlg \circ \Calc.$

  We can extend this naive type system with fixed points and with modal operators that denote sets of terms that all exhibit a particular behavior; for an example of the latter, the arrow type $A \Rightarrow B$ in the $\lambda$-calculus is the type of terms that when applied to a term of type $A$ eventually rewrite to a term of type $B$.  Other formulae can capture eventual properties like
\begin{itemize}
  \item Authority: it is the case that if my banking service eventually evolves to a state in which Alice has the ability to withdraw money from my account, then the banking service both sent an email to my address requesting confirmation and received the confirmation email in response.
  \item Confinement: this process will only ever send a messages on channels in this set.
  \item Liveness: this process will always eventually handle another message.
  \item Termination: this computation will always halt.
  \item Structure: at the end of sorting, the data will be ordered.
\end{itemize}

The slogan ``logic via distributive law'' is more apt than might first
be supposed, however. At heart, the algorithm says that logical
formula can be thought of as describing collections of terms in terms
of terms built over collections. Since a very broad class of
collections are captured via the notion of monad, the distributive
law, together with the monad laws are iteratively applied to instances
of the latter to arrive at instances of the former. In this sense
$\BoolAlg$ is just one notion of ``collection'', namely that of
$\Set$. There are other notions of collection, such as list, bag,
tree, all of which are captured as monads and for which we could use a
distributive law to provide a semantics for a notion of formula that
has utility and considerable expressive power. Thus, for example, if
the notion of collection were \emph{sets of sequences}, then the
semantics generated by our algorithm would interpret formulae as sets
of sequences of terms, instead of sets of terms, and the term language
extracted for the collection monad would be the familiar operators of
classical linear logic and this would correspond to the well-known
fact that quantales provide full and faithfull models of the
\emph{formulae} of linear logic. The higher categorical machinery
provides the right setting in which to express this idea in it's full
generality. In this paper, however, we focus the exegesis of these
results on $\BoolAlg$ to conserve space, and highlight the core ideas.

\section{Related work}

There is a very large body of work on Lawvere theories and their
generalizations, and we cannot give a proper review here; instead we
list a few key papers with many references.  Hyland and Power
\cite{DBLP:journals/entcs/HylandP07} reviewed the history of both
Lawvere theories and monads as an approach to universal algebra, how
monads came to the fore, Plotkin's work, and the renewed interest in
Lawvere theories in a computer science setting.  Trimble
\cite{Trimble} constructed multisorted Lawvere theories.  Lack and
Rosick\'y \cite{DBLP:journals/acs/LackR11} unified the work of several
earlier authors by considering models in categories other than Set,
extending the notion of Lawvere theory to enriched categories, and
using all limits rather than just products all at once.

Our results can be seen as a natural consequence of Abramsky's
programme, laid out in ``Domain theory in Logical Form''
\cite{DBLP:journals/apal/Abramsky91}, and expanded in ``Proofs as
Processes'' \cite{Abramsky:1992:PP:194588.194591}.  Moreover, they can
be seen as bringing the seminal work of Caires \cite{Caires} under the
rubric of that programme.

\section{A 1-categorical example}

Here we return in slightly greater detail to the example of the logic for the language of monoids.  Because we are working with categories instead of 2-categories, the ``virtual machine'' is particularly bland; it only has terms (the elements of the monoid), not rewrites.

Let FinSet be a skeleton of the category of finite sets.  The Lawvere
theory of a computational calculus is a category Th(Calc) with finite
products equipped with an identity-on-objects functor $\theta\maps
\FinSet^\op \to \mathrm{Th(Calc)}.$  Because the objects of FinSet are
coproducts of the one-element set, the objects of category Th(Calc)
are therefore products of a generating object we will write as $S,$
for ``sort''.  The morphisms of Th(Calc) are generated from a set of
morphisms from finite powers of $S$ to $S$ by products and
composition, so Th(Calc) may be presented by

\begin{itemize}
  \item a sort $S$,
  \item a set of term constructors $f_i\maps S^{n_i} \to S,$ where $i$ ranges over some index set $I$ and $n_i \in \NN,$
  \item and a set of equations between term constructors.
\end{itemize}
We say the arity of $f_i$ is $n_i.$

A product-preserving functor from Th(Calc) to Set picks out a set and equips it with structure maps satisfying the equations.  The category Prod(Th(Calc), Set) of product-preserving functors and natural transformations between them is equivalent to the category Calc of calculi and calculus homomorphisms.  There is a forgetful functor ${U\maps \Calc \to \Set}$ that forgets the extra structure.  The functor $U$ has a left adjoint ${L\maps \Set \to \Calc}$ that picks out the free calculus on a set of base terms.  The monad $\Calc = UL\maps \Set \to \Set$ picks out the underlying set $ULX$ of the free calculus $LX$ on a set $X$.  Lawvere theories are in bijection with finitary monads; the qualifier ``finitary'' means that each term constructor has a finite arity.

Here is a presentation of the ``Lawvere theory of a monoid'' Th(Mon):
\begin{center}
  \begin{longtable}{|p{0.3\linewidth}|p{0.7\linewidth}|}
    \hline
      Sorts:
      \begin{itemize}
        \item $S$
      \end{itemize}\bigskip
      Term constructors:
      \begin{itemize}
        \item $\cdot\maps S^2 \to S$
        \item $e\maps 1 \to S$
      \end{itemize}
    &
      Equations:
      \begin{itemize}
        \item $\cdot \circ (S \times \cdot) = \cdot \circ (\cdot \times S)$ (associativity)
        \item $\cdot \circ (e \times S) \circ \leftu^{-1} = S$ (left unit)
        \item $\cdot \circ (S \times e) \circ \rightu^{-1} = S$ (right unit)        
      \end{itemize}\\
    \hline
  \end{longtable}
\end{center}
where ${\leftu\maps 1 \times S \stackrel{\sim}{\to} S}$ and ${\rightu\maps S \times 1 \stackrel{\sim}{\to} S}$ are the canonical isomorphisms.

An implementation of this specification is a product-preserving functor from Th(Mon) to Set; such a functor will assign a set of values to the sort and functions to the term constructors such that the equations are satisfied, {\em i.e.} it will pick out a monoid.  The category of product-preserving functors from Th(Mon) to Set and natural transformations between them is equivalent to the category of monoids and monoid homomorphisms.  There is a forgetful functor $U\maps \Mon \to \Set$ that forgets the multiplication and unit, and outputs the underlying set of elements.  The functor $U$ has a left adjoint $L\maps \Set \to \Mon$ that outputs the free monoid on a set.  The composite functor $\Mon = UL\maps \Set \to \Set$ is the corresponding monad.

All our formulae about monoids will denote subsets of the elements of the monoid; we use Th(BoolAlg) to describe them.  \begin{center}
  \begin{longtable}{|p{0.3\linewidth}|p{0.7\linewidth}|}
    \hline
    Sorts:
    \begin{itemize}
      \item $S$
    \end{itemize}
    Term constructors:
    \begin{itemize}
      \item $\land, \lor\maps S^2 \to S$
      \item $\top, \bot\maps 1 \to S$
      \item $\neg\maps S \to S$
    \end{itemize}
    &
    Equations:
    \begin{itemize}
      \item \raggedright associativity, commutativity and unit laws for $\land$ and $\lor$
      \item distributivity of $\land$ over $\lor$
      \item involution for $\neg$
      \item de Morgan's laws
    \end{itemize}\\
    \hline
  \end{longtable}
\end{center}
From this theory, we can derive the monad BoolAlg for the free Boolean algebra on a set.

The sum of the two theories above is a new theory Th(Form) = Th(Mon) + Th(BoolAlg) whose terms are our formulae.  Th(Form) is presented by identifying the sorts and taking the union of the term constructors and the union of the equations.
\begin{comment}
  \begin{longtable}{|p{0.3\linewidth}|p{0.7\linewidth}|}
    \hline
    Sorts:
    \begin{itemize}
      \item $S$
    \end{itemize}
    Term constructors:
    \begin{itemize}
      \item $\cdot\maps S^2 \to S$
      \item $e\maps 1 \to S$
      \item $\land, \lor\maps S^2 \to S$
      \item $\top, \bot\maps 1 \to S$
      \item $\neg\maps S \to S$
    \end{itemize}
    &
    Equations:
    \begin{itemize}
      \item associativity and unit laws for $\cdot$
      \item \raggedright associativity, commutativity, and unit laws for $\land$ and $\lor$
      \item involution for $\neg$
      \item de Morgan's laws
    \end{itemize}\\
    \hline
  \end{longtable}
\end{comment}

The process of deriving a monad from a theory preserves sums.  Since the sum of two monads is the free product of the two, a general formula will be a term in an alternating composition of the two monads.  For example, suppose that $a, b, c, d \in X;$ then one formula is
\[ (({a}\lor{(b \cdot d)}) \cdot ({c}\lor{d})), \]
which is a term in Mon(BoolAlg(Mon($X$))).  The interpretation of this formula should be the set of monoid elements
\[ \{ ac, ad, bdc, bdd \}, \]
or in other words, the term
\[ (a \cdot c) \lor (a \cdot d) \lor ((b \cdot d) \cdot c) \lor ((b \cdot d) \cdot d) \]
in BoolAlg(Mon(X)).

In order to move all the uses of Mon to the right of the uses of BoolAlg in the alternating composition, we need a distributive law natural transformation
\[ \delta\maps \Mon \circ \BoolAlg \Rightarrow \BoolAlg \circ \Mon. \]
Given $\delta$ and the monad units and multiplications, we can define an interpretation natural transformation
\[ \interp{-}\maps \Form\Rightarrow \BoolAlg \circ \Mon \]
in the obvious way.  Below, we write subsets of Mon($X$) using set notation:
\begin{center}
  \begin{longtable}{|rl|rl|}
    \hline
    $\interp{\top}_X$ &= $\Mon(X)$ &
    $\interp{\bot}_X$ &= $\emptyset$ \\
    $\interp{{A}\lor{B}}_X$ &= $\interp{A}_X \cup \interp{B}_X$ &
    $\interp{{A}\land{B}}_X$ &= $\interp{A}_X \cap \interp{B}_X$\\
    $\interp{\neg A}_X$ &= $\Mon(X) - \interp{A}_X$ &
    $\interp{{A} \cdot {B}}_X$ &= $\Mon(\cdot)(\interp{A}_X \times \interp{B}_X)$\\
    $\interp{e}_X$ &= $\{ e\}$ &
    $\interp{x \in X}_X$ &= $\{ x\}$\\
    \hline
  \end{longtable}
\end{center}

Even in this simple example, we have nontrivial formulae; for example,
\[ prime = \neg e \land \neg(\neg e \cdot \neg e) \]
is a 1-line formula for primality.  For the monoid of natural numbers under multiplication, this formula says a number is prime if it is neither 1 nor has a nontrivial factorization.  It is easy to verify that $\interp{prime}_X = X.$

\section{Moving to 2-categories}

In a computational context, when a data structure has some symmetry we do not care about, we often test two instances of the structure for equality by computing a normal form for each instance and then comparing the normal forms.  Monoids are associative, but we can imagine storing words of a monoid as binary trees internally and then comparing them by computing a normal form.  For the case of a normal form for the trees, we can eliminate 1 in a product by rewriting $(1 \cdot x)$ and $(x \cdot 1)$ to $x$, and we can shift all the parentheses to the right by rewriting $((x \cdot y) \cdot z)$ to $(x \cdot (y \cdot z)).$

The Lawvere 2-theory Th(Mon) is much the same as the 1-theory, except for the weakening of some equations to rewrites and the addition of new equations between the rewrites.

\begin{center}
  \begin{longtable}{|p{0.3\linewidth}|p{0.7\linewidth}|}
    \hline
    Sorts:
    \begin{itemize}
      \item $S$
    \end{itemize}
    Term constructors:
    \begin{itemize}
      \item $\cdot\maps S^2 \to S$
      \item $e\maps 1 \to S$
    \end{itemize}
    &
    Rewrites:
    \begin{itemize}
      \item $a\maps \cdot \circ (S \times \cdot) \Rightarrow \cdot \circ (\cdot \times S)$
      \item $l\maps \cdot \circ (e \times S) \circ \leftu^{-1} \Rightarrow S$
      \item $r\maps \cdot \circ (S \times e) \circ \rightu^{-1} \Rightarrow S$
    \end{itemize}
    Equations:
    \begin{itemize}
      \item $a \circ a = (S \times a) \circ a \circ (a \times S)$ (pentagon equation)
      \item $r \times S = (S \times l) \circ a$ (triangle equation)
    \end{itemize}\\
    \hline
  \end{longtable}
\end{center}
From Th(Mon) we can derive a 2-monad Mon that essentially produces the free monoid on a set, but keeps track of the internal representation of the monoid---a binary tree---and accounts for the work needed to convert the internal representation to its normal form.  From this perspective, we can think of the trees as programs for a simple computational calculus and the process of normalization as the execution of the program.  Later in the paper, we will examine the case of the SKI combinator calculus, a Turing-complete programming language that was a predecessor to the $\lambda$-calculus; it, too, uses binary trees as programs, and execution of the program is a normalization process.

When we add the 2-monad BoolAlg to Mon, we get formulae like $(1 \cdot \top)$ denoting the set of trees whose leftmost child is the identity; because Mon is a 2-monad, we also get proofs like 
\[ ((r \circ l) \lor y)\maps ((e \cdot (x \cdot e)) \lor y) \to (x \lor y) \]
whose interpretations are homomorphisms of Boolean algebras.

The formulae in these examples have been propositions about the structure of terms.  Later in the paper, we will also show how to add modal operators that are propositions about the behavior of terms; the arrow type constructor from the $\lambda$-calculus is a prominent example.

\section{Multisorted Lawvere theories}
In the motivation section, each theory had only one sort; practical theories for virtual machines are usually multisorted.  Given a finite set of sorts $\Sigma,$ the category $\FinSet/\Sigma$ is the category whose objects are pairs $(S, s\maps S\to \Sigma),$ where $S$ is a finite set, and whose morphisms are functions $f\maps S \to S'$ such that the relevant triangle commutes.

A multisorted Lawvere theory is a category Th(Calc) with finite products equipped with an identity-on-objects functor ${\theta\maps (\FinSet/\Sigma)^\op \to \mathrm{Th(Calc)}.}$  The category Prod(Th(Calc), $\Set^\Sigma$) is equivalent to the category Calc of calculi and calculus homomorphisms.  There is a forgetful functor ${U\maps \Calc \to \Set^\Sigma}$ with a left adjoint ${L\maps \Set^\Sigma \to \Calc}$ that picks out the free calculus on a $|\Sigma|$-tuple of sets.  The monad ${UL\maps \Set^\Sigma \to \Set^\Sigma}$ picks out the underlying $|\Sigma|$-tuple of sets $ULX$ of the free calculus $LX$ on a $|\Sigma|$-tuple of sets $X$.

An example of a multisorted Lawvere theory is that of a group action on a set, which involves a choice of both a group $G$ and a set $V$ to act on.  The presentation of Th(GrpAct) has a pair of sorts $(G, V)$, all the term constructors and equations as the theory of a group (where we replace $S$ by $G$), together with a new term constructor
\begin{itemize}
  \item $a\maps G \times V \to V$
\end{itemize}
and equations
\begin{itemize}
  \item $a \circ (e \times V) \circ \leftu^{-1}_V = V$ (identity action)
  \item $a \circ (m \times V) = a \circ (G \times a)$ (compatibility).
\end{itemize}

Another example is the theory of a directed graph, with one sort for vertices and another for edges, with source and target maps for term constructors.

\section {Lawvere 2-theories}
In this paper, the multisorted Lawvere 2-theory of a calculus is a 2-category Th(Calc) with strict finite products (that is, its underlying category has products) equipped with an identity-on-objects functor $\theta\maps (\FinSet/\Sigma)^\op \to \mathrm{Th(Calc)},$ where we promote $(\FinSet/\Sigma)$ to a 2-category by adding identity 2-morphisms to every 1-morphism.  As noted in the related work section, other authors have considered much more powerful notions of 2-theory, but we will, for the most part, not need the extra features; however, see the notion of ``lambda theory'' below.  Our notion of a multisorted Lawvere 2-theory may be presented by a finite set of sorts, a set of term constructors with finite arity, a set of rewrites, and a set of equations between rewrites.  

Our models of multisorted Lawvere 2-theories are functors into $\Cat^\Sigma$ that preserve products up to isomorphism, not merely up to equivalence; that is, the 2-functor has an underlying functor that preserves products.  As with 1-theories, the 2-category of product-preserving functors from Th(Calc) to $\Cat^\Sigma$, natural transformations, and modifications is equivalent to the 2-category of calculi, calculus homomorphisms, and calculus transformations.

As mentioned above, the SKI combinator calculus is a Turing-complete language; it was invented by Sch\"onfinkel \cite{Schonfinkel} and Curry \cite{Curry} in the 1920s as a way to clarify the role of quantified variables in logic, essentially by eliminating them.  The single-sorted Lawvere 2-theory Th(SKI) has a presentation
\begin{center}
  \begin{longtable}{|p{0.3\linewidth}|p{0.7\linewidth}|}
    \hline
    Sorts:
    \begin{itemize}
      \item $T$
    \end{itemize}
    Term constructors:
    \begin{itemize}
      \item $S, K, I\maps 1 \to T$
      \item $(-\;-)\maps T^2 \to T,$
    \end{itemize}
    &
    Rewrites:
    \begin{itemize}
      \item $\forall x,y,z \in T, \quad \sigma \maps (((S\; x)\; y)\; z) \Rightarrow ((x\; z)\; (y\; z))$
      \item $\forall y,z \in T, \quad \kappa \maps ((K\; y)\; z) \Rightarrow y$
      \item $\forall z \in T, \quad \iota \maps (I\; z) \Rightarrow z$
    \end{itemize}
    No equations.\\
    \hline
  \end{longtable}
\end{center}
In this context, the Church-Rosser theorem for the SKI calculus says that any two terminating rewrites out of an SKI term have the same codomain.  We do not usually want to impose equality on the rewrites, since they can differ greatly in computational complexity.  For example, suppose that we have the term $((K\; I)\; x),$ where $x$ is some term that takes a long time to reduce to its normal form; a rewrite that reduces $x$ first and then uses $\kappa$ takes much longer than just doing $\kappa$ first, though both rewrites begin and end at the same term.

The free model of Th(SKI) on a category takes its objects as terms and its morphisms as rewrites, then freely adjoins $S,K,$ and $I$ and all applications of one object to another, as well as new morphisms generated by $\sigma,\kappa,$ and $\iota.$  The free model on the empty category will contain only terms and rewrites from the SKI calculus.

Just as Lawvere theories capture the notion of a set equipped with functions satisfying equations, 2-Lawvere theories capture the notion of a category equipped with functors and natural transformations satisfying equations.  The former describes what might be called data structures, while the latter captures both the state of a system and the processes that update that state.  In the next section, we introduce a language for talking about sets of states and updates that satisfy some criterion, terms and rewrites that share some property.

\section{Categories of formulae and proofs}

Any 1-theory can be promoted to a 2-theory by turning each equation into a rewrite, then adding an equation asserting that the rewrite is equal to the identity rewrite.  Therefore as before, given a single-sorted Lawvere 2-theory Th(Calc), we get a 2-theory of formulae Th(Form) by adding the 2-theory of Boolean algebras Th(BoolAlg).  The models of Th(Form) are categories of formulae and proofs.

For multisorted Lawvere 1- and 2-theories, there is no canonical choice of sorts to identify between the theory of the calculus and the theory of Boolean algebras; one may choose to add one or more different copies of Th(BoolAlg) and identify each single sort with different sorts in the theory of the calculus.  For example, given the theory of a group action, one could add a copy of Th(BoolAlg) for both $G$ and $V$ and write formulae involving subsets of the group and of the set it acts on.  Later in the paper, we will see how we can use formulae involving names to create namespaces that enforce confinement on processes in the $\pi$-calculus.

\section{Interpretation}

When describing the semantics of a logic, questions of soundness and
completeness are natural.  The whole point of our approach is that it
is correct by construction.  Soundness and completeness for the
modal-free fragment follow directly from the definitions; soundness,
in particular, is a direct consequence of the way realizability is
incorporated into the approach.  For the formulae, the intuitions
underlying the completeness argument amount to the fact that the term
language for the logical operators associated with the collections are
extractions of a term language for the collection monad.  Thus, they
are in perfect correspondence with the model.  In the case of $\Set$,
this is equivalent to the well-known fact that boolean algebras are
effectively realized by the powerset lattice.

Likewise, in the case of $\Set$, the semantics that maps formulae to
the composition of the collection monad with the monad for the term
language supporting computation is nothing more than a pointwise
lifting of the powerset semantics.  The categorical presentation
provides an appropriately abstracted form of this simple
observation. In particular, interpretation is a natural transformation
between 2-monads on Cat; it assigns to each category $X$ a functor
\[\interp{-}_X\maps (\BoolAlg + \Calc)(X) \to (\BoolAlg \circ \Calc)(X).\]
Functoriality means that the interpretation is sound; if the functor is full, the interpretation is complete. Since we define interpretation in terms of the monad laws and BoolAlg + Calc is the free product on the two monads, $\interp{-}_X$ is the identity when restricted to $(\BoolAlg \circ \Calc)(X)$ and thus full.

Notice we have said nothing about the faithfulness of this functor. In
particular, if we consider rewrites as equalities then proofs (and
therefore homsets) collapse, recovering the usual situation of posets as models
of classical logic. On the other hand, when they are not treated as
equalities, we are able to distinguish proofs, even in the classical
setting. There is much, much more to say about this and other
phenomena associated with Joyal's lemma \cite{0910.2401}, but we defer the discussion
to another paper.

\section{Modal operators}

So far, all the examples of formulae have dealt with the structure of the term.  Far more interesting are sets of terms that all share some behavior.  For example, in the SKI calculus, we want to add the idea of an arrow type to our formulae, and have the interpretation of $A \Rightarrow B$ be the subset of terms $t$ such that given a term $u\in\interp{A}$, the term $(t\; u)$ eventually evolves to a term $v\in\interp{B}.$

This notion of eventuality is an example of a modal operator.  Possibility is another, where possibility is to eventuality as ``there exists'' is to ``for all''.  In the SKI calculus, they coincide due to the fact that the calculus is confluent, but they differ in the general case.  Many interesting properties can be stated in terms of eventual and possible states as noted earlier.

To add a modal operator to a formula language, we first add it formally as a term constructor to the term theory, then add the collection theory as before.  

To interpret the modal operator, we first interpret the formulae as above, then interpret modal terms as collections of terms, then use the join from the collection monad.

\subsection{Example: Arrow types in the SKI calculus}

The Lawvere 2-theory of the arrow type Th(Arrow) has one term constructor, no rewrites, and no equations.  The lack of rewrites and equations are because the arrow is a purely formal type, and it is only in our choice of semantics that it acquires its customary interpretation.

In order to avoid notational confusion between the arrow type constructor and 2-morphisms, we will use a triple-arrow in the theory.
\begin{center}
  \begin{longtable}{|p{0.3\linewidth}|p{0.7\linewidth}|}
    \hline
    Sorts:
    \begin{itemize}
      \item $T$
    \end{itemize}
    Term constructors:
    \begin{itemize}
      \item $\Rrightarrow \maps T^2 \to T$
    \end{itemize}
    &
    No rewrites; no equations.\\
    \hline
  \end{longtable}
\end{center}

We get the theory of SKI with arrow Th(SKIArr) by adding Th(SKI) to Th(Arrow), and we get the theory of formulae Th(Form) by adding Th(SKIArr) to Th(BoolAlg).

To interpret terms from Th(Form), we compose the interpretation natural transformation above with another natural transformation $\alpha\maps \mathrm{SKIArr} \Rightarrow \mathrm{BoolAlg} \circ \mathrm{SKI},$ followed by the join from the collection monad.  In particular, 
\[ \alpha(u \Rrightarrow v) = \{ t \;|\; \exists \rho\maps (t\; u) \Rightarrow v \}. \]

In Lambek's 1980 paper \cite{Lambek} on the denotational semantics of the $\lambda$-calculus, he defined a category whose objects were types and whose morphisms were equivalence classes of lambda terms with one free variable.  In a future paper, we will show how Mellies and Zeilberger's approach to type refinement lets us recapitulate Lambek's construction and extend the arrow to a profunctor.

\subsection{Modal operators parametric in a term constructor}

A term context is a term with a ``hole'' that can be filled by some other term.  Given a Lawvere theory with a sort $S$ for terms, one can derive a new theory of term contexts by replacing each term constructor $f\maps S^n \to S$ with $n$ term constructors $f_i\maps S^{n-1} \to S$ where $1 \le i \le n$.  We think of $f_i$ as being $f$ with the $i$th input being a hole.  Equivalently, if we have a theory with coproducts, we can replace each occurrence of $S$ on the left-hand side of a term constructor with $S+1,$ where the new point represents the hole.

The arrow type is a special case of a more general modal operator parametrized by a two-hole term context $C.$  We denote the operator itself with angle brackets as a reminder of the diamond ``possibly'' modality: $u \langle C \rangle v.$  The interpretation is similar to that of the arrow:
\[ \interp{u \langle C\rangle v} = \{ t \;|\; \exists \rho\maps C[t, u] \Rightarrow v \}; \]
that is, $u \langle C\rangle v$ denotes those terms $t$ that when put
in the context $u$ may possibly evolve to $v.$

This definition is inspired by the work of Sewell, Leifer, and Milner, in which they formalize the notion of a context-labelled transition \cite{DBLP:conf/concur/LeiferM00}.  Given the now standard Hennessy-Milner interpretation of modal logics which interpret possibility in terms of actions labelling transitions, when Sewell {\em et al.} gave a notion of context-labelled transitions, it was only natural
to consider modal operators based on contexts.

\section{Recursion}

Suppose that we have a single-sorted Lawvere 2-theory of a calculus with sort $S$.  We add recursion to our formulae by introducing a new sort $V$ for type variables and new term constructors 
\begin{itemize}
  \item $- \maps V \to S$ to let us use type variables in formulae and
  \item $\mu\maps V\times S \to S$ to express fixed points.
\end{itemize}

We interpret terms of the form $\mu X.P[X]$ as the greatest fixed point of $P[X].$

\subsection{The reflective higher-order $\pi$-calculus}

Lawvere theories are limited in that they only talk about products of sorts.  A lambda theory is a generalization of a Lawvere theory that has the ability to talk about function sorts like $A \Rightarrow B$ and sums like $A+B$ or $1 + A + A^2 + \cdots,$ more commonly denoted with the Kleene star $A^*.$  One can think of lambda theories as having access to a ``library'' that takes care of the details of bound variables and substitution for us so we do not have to implement all that machinery.  Much of the work on nominal logics has been about exactly this factorization \cite{GabbayMJ:picfm}.  A lambda theory Th(Calc) is a bicartesian closed 2-category equipped with an identity-on-objects functor from $(\FinSet/\Sigma)^{\op}$ to Th(Calc).  Models of Th(Calc) are functors to Cat that preserve all the structure.

The $\pi$-calculus was invented in the early 1990s by Robin Milner as a model of networks of processes with a dynamically changing topology; two processes initially unaware of each other can be introduced by a third process.  The reflective higher order $\pi$-calculus uses quoted processes as names; the term constructors for quote and eval replace the more traditional nu and replicate constructors.  We also add a ``comm'' term to restrict the contexts in which reduction can occur \cite{DBLP:journals/corr/StayM15}.

Here is a presentation of the multisorted lambda theory Th(RHOpi) for the reflective higher-order $\pi$-calculus:
\begin{center}
  \begin{longtable}{|p{0.3\linewidth}|p{0.7\linewidth}|}
    \hline
    Sorts:
    \begin{itemize}
      \item $N$ for names
      \item $P$ for processes
    \end{itemize}\bigskip
    Term constructors:
    \begin{itemize}
      \item $\send\maps N \times P^* \to P$
      \item \raggedright $\recv\maps N \times (N^* \Rightarrow P) \to P$
      \item $|\maps P^2 \to P$
      \item $0\maps 1 \to P$
      \item $\comm\maps 1 \to P$
      \item $\quote{-}\maps P \to N$
      \item $\deref{-}\maps N \to P$
    \end{itemize}
    &
    Rewrites:
    \begin{itemize}
      \item $\alpha\maps (p_1 | p_2) | p_3 \Rightarrow p_1 | (p_2 | p_3)$
      \item $\beta\maps p_1 | p_2 \Rightarrow p_2 | p_1$
      \item $\iota\maps 0 | p \Rightarrow p$
      \item \raggedright $\chi\maps \send(x, p_1, \ldots, p_n) \;|\; \recv(x, q) \;|\; \comm \Rightarrow q(\quote{p_1}, \ldots, \quote{p_n}) \;|\; \comm$
      \item $\epsilon\maps \deref{\quote{p}} \Rightarrow p$
    \end{itemize}
    Equations:
    \begin{itemize}
      \item \tshortstack[l]{$\alpha = P^3, \beta = P^2, \iota = P$ ($|$ and 0 form a \\ commutative monoid)}
      \item \tshortstack[l]{$\epsilon = P$ (evaluating a quoted process is the \\ same as the process itself)}
    \end{itemize}\\
    \hline
  \end{longtable}
\end{center}
The simplest RHOpi processes are 0, the ``do nothing'' process; and comm, a ``catalyst'' process that enables communication on a channel.  The only rewrite that is not an identity is $\chi,$ the communication event.  The $\chi$ rewrite is neither confluent nor deterministic.  For example, we can model contention for resources with the term
\[ \recv(x, P)\;|\;\recv(x, Q)\;|\;\send(x,R)\;|\;\comm \]
which has two rewrites out of it, one where the continuation $P$ is invoked on the name $\quote{R}$ and the other where the continuation $Q$ is invoked on it.  We can model message arrival order nondeterminism with the term
\[ \send(x, P)\;|\;\send(x, Q)\;|\;\recv(x,R)\;|\;\comm \]
which has two similar rewrites out of it, one where the continuation $P$ is invoked on the name $\quote{R}$ and one where $P$ is invoked on the name $\quote{Q}$.

In the theory above, comm is preserved by the rewrites; one can think of each comm instance as representing a processor.  An alternative would be to consume comm in the $\chi$ rewrite; then comm would track clock ticks; an application of a consumable comm is the formal verification of billing code for tracking compute resources.

Replication of processes, and therefore general recursion, can be encoded \cite{DBLP:journals/entcs/MeredithR05} via
\[D(x) = \recv(x, y\mapsto \send(x, \deref{y}) | \deref{y})\]
\[!P = \send(x, D(x) | P) | D(x).\]

Caires' \cite{Caires} operator $\triangleright$ for rely-guarantee properties, the adjunct to $|,$ is an instance of our generic modal operator, where $u \triangleright v = u \langle - | - \rangle v:$
\[ \interp{u \triangleright v} = \{ t \;|\; \exists \rho\maps (t\;|\;u) \Rightarrow v\} \]

The logic is generated as above, adding modal operators and recursion to the term language, then adding the Boolean algebra; it is equivalent to the logic we generated by hand in \cite{DBLP:conf/tgc/MeredithR05}.  As an example of a formula in this logic, here is a one-line formula for a process that will always be able to handle another message:
\[ \mu X. \recv(\top, x \mapsto X)\;|\;\top. \]
It says that the process must factor into a piece that is waiting for a message on some channel, and when it receives that message, the continuation will have the same form.

\subsubsection{Namespaces}

When we add a copy of BoolAlg to RHOpi for both $N$ and $P,$ we can write down formulae that talk about sets of names.  The formula $\top$ denotes the set of all names; the formula $\quote{\top}$ denotes the set of names that are quoted processes.  These two sets differ when we have a nonempty set of generating names.

An application of namespace logic is a formula for a ``firewall'': any process satisfying this formula receives no messages except on a name in $\quote{\phi}.$
\[ \mu X. \recv(\quote{\phi}, x \mapsto (X \lor 0)\;|\;\neg\recv(\quote{\neg \phi}, \top))\;|\;\neg\recv(\quote{\neg \phi}, \top) \]
It is, perhaps, surprising that this can be accomplished with a compile-time check, since we usually think of firewalls as a dynamic check.  An application of namespaces to statically proving security properties of code in an object capabilities language can be found in \cite{DBLP:journals/corr/MeredithSD13}.

\section{Conclusion and future work}
We have presented an algorithm for generating a logic from three data:
a 2-Lawvere theory describing a notion of computation, a monad for
describing a notion of collection, and a distributive law. As
mentioned in the introduction, the picture emerging from this view of
logic is that logically useful collections of individuals look like
nothing so much as the individuals that represent the formulae that
collect them.  The range of applications of this algorithm is quite
broad.  Any of the languages given semantics in the K Framework,
including C/C++, Java, Javascript, and many others can automatically
be equipped with logics that serve as a basis for of program
specification and automatically checked verification, and
simultaneously as a basis equipping these languages with new and
powerful type systems.  This considerably lowers the barrier to
embuing popular untyped languages with type systems.  Likewise, the
algorithmic nature of the approach also serves as an aid in future
language design.  For example, Synereo is designing a language for the
smart contracts on the blockchain based on the
$\mathsf{RHO}$-calculus \cite{rholang}.  Equipping this language with a type system
amounts to nothing more than deciding on the type of collection one
wants to collect type inhabitants in.

\bibliographystyle{amsplain}
\bibliography{ladl}
\end{document}